# Effects of Zeeman spin splitting on the modular symmetry in the quantum Hall effect

D. R. Hang[1,2], R. B. Dunford[3,4], Gil-Ho Kim[5], H. D. Yeh[6,7], C. F. Huang[6,7], D. A. Ritchie[3], I. Farrer[3], Y. W. Zhang[2], C.-T. Liang[3,7], and Y. H. Chang[7]

[1]*Department of Materials Science and Optoelectronic Engineering, National Sun Yat-sen University, Kaohsiung 804, Taiwan*

[2]*Institute of Materials Science and Engineering, National Sun Yat-sen University, Kaohsiung 804, Taiwan, R.O.C.*

[3]*Cavendish Laboratory, Madingley Road, Cambridge CB3 0HE, United Kingdom*

[4]*Division of Information and Communication Sciences, Macquarie University, NSW 2109, Australia*

[5]*Department of Electronic and Electrical Engineering, Sungkyunkwan University, Suwon 440-746, Korea*

[6]*National Measurement Laboratory, Center for Measurement Standards, Industrial Technology Research Institute, Hsinchu 300, Taiwan*

[7]*Department of Physics, National Taiwan University, Taipei 106, Taiwan, R.O.C.*

Magnetic-field-induced phase transitions in the integer quantum Hall effect are studied under the formation of paired Landau bands arising from Zeeman spin splitting. By investigating features of modular symmetry, we showed that modifications to the particle-hole transformation should be considered under the coupling between the paired Landau bands. Our study indicates that such a transformation should be modified either when the Zeeman gap is much smaller than the cyclotron gap, or when these two gaps are comparable.

Two types of magnetic-field-induced phase transitions, plateau-plateau and insulator-quantum Hall (I-QH) transitions, can be observed in the quantum Hall effect (QHE) as we applied a magnetic field $B$ perpendicular to a two-dimensional (2D) system.[1-3] Based on the law of corresponding states[1], universal features of the $\Gamma_0(2)$ modular symmetry such as the semicircle law and universal critical conductivities can be found in all magnetic-field-induced phase transitions.[2] In such a law, however, effects due to the Zeeman spin splitting are ignored. Dolan suggested that the spin splitting can reduce the modular symmetry from $\Gamma_0(2)$ to $\Gamma(2)$, under which the universality of the critical conductivities can become invalid while the trace $\sigma_{xx}(\sigma_{xy})$ follows the semicircle law.[2] Here $\sigma_{xx}$ and $\sigma_{xy}$ are the longitudinal and Hall conductivities. The deviation of the critical point can be explained by the generalization of the particle-hole transformation under $\Gamma(2)$ symmetry.[2] In most 2D systems, as shown in Fig. 1 (a), the cyclotron splitting is much larger than the spin splitting. The two Landau bands separated by the spin gap are paired, and the coupling between the paired bands can reduce the modular symmetry to $\Gamma(2)$. In addition, recent studies suggest that the (generalized) particle-hole transformation may need to be modified as well when the gap inducing QH state is small.[4] Since such a transformation is important to the semicircle law[2,5], the deviation of such a law is also expected under a small spin gap.

In this paper we investigate three different 2D systems to study effects of Zeeman spin splitting in the integer quantum Hall effect (IQHE). We benefit from the fact that these samples have been examined in Refs. 3, 6 and 7. Table I summarizes the systems under study and the corresponding transitions. For convenience, we denote a QH state by its filling factor ν and the insulating state by the number "0". In addition, a magnetic-field-induced transition is characterized by its adjacent two states.[1-3]

Figure 2 shows the traces of the longitudinal and Hall resistivities $\rho_{xx}$ and $\rho_{xy}$ of sample B at the gate voltage $V_g = -0.26$ V. [6] At $B=0$, the sample behaves as an insulator and $\rho_{xx}$ increases with decreasing the temperature $T$. With increasing $B$, the sample undergoes a 2-0 transition at the point $C_2$ to enter the quantum Hall liquid from the low-field insulator. The spin splitting is resolved at higher $B$, where the 2-1 and 1-0 spin-resolved transitions are observed. Thus in sample B, we can probe the 2-0, 2-1, and 1-0 transitions. In Fig. 3, the red solid line and dash dot line are the traces of $\sigma_{xx}(\sigma_{xy})$ in the 2-0 and 2-1 transitions of sample B when the temperature $T=$ 50 mK. In the 2-0 transition, $\sigma_{xx}(\sigma_{xy})$ is along the expected semicircle, the green dashed line, when $\sigma_{xy} > e^2/h$ because of the validity of the semicircle law at the side of QH liquid.[8] However, the trace $\sigma_{xx}(\sigma_{xy})$ is not along the semicircle in the 2-1 transition, which indicates that the semicircle law is not strictly followed as the spin splitting is just resolved. On the other hand, as shown in Fig. 4, a well-defined $T$-independent point can be identified in $\sigma_{xy}$ in the 2-1 transition. Such a point is the critical point, and we can see that the critical Hall conductivity is close to the universal value 1.5 $e^2/h$. Therefore, the universality of critical Hall conductivity can survive under the invalidity of the semicircle law when the spin splitting is just resolved. In fact, it is expected that universalities are very robust in $\sigma_{xy}$ since $\sigma_{xy}$ is an important quantity to construct the phase diagram of QHE.[4,9] However, it is also shown by our group[3] that the semicircle law holds well in sample A with the invalidity of universal critical conductivities when the spin splitting is resolved but is still small. In Fig. 3, the critical points of the 2-1 and 1-0 transitions of sample A deviate from the top position of the semicircle, which indicates the invalidity of universal critical conductivities.[2] The deviation in the 2-1 transition indicates $\Gamma(2)$ symmetry.[2]

To further study effects due to Zeeman spin splitting, we investigate the transitions in sample C. The sample is of a p-type Si/SiGe heterostructure, where the spin splitting is comparable to the cyclotron energy.[7] As shown in Fig. 1 (b), the Landau bands also appear in pairs except the lowest Landau band in such heterostructures. When the paired Landau bands are unresolved, only QH states of odd filling factors can be observed. In sample C, the $\nu = 2$ state is the first well-resolved QH state of the even filling factor, and thus the 3-2 and 2-1 transitions are suitable to study effects due to paired Landau bands. In Fig. 3, the blue dotted line is the trace $\sigma_{xx}(\sigma_{xy})$ in the 3-2 transition at $T = 70$ mK, and the open diamond (C: 3-2) denotes the corresponding critical point located at (2.76 $e^2/h$, 0.46 $e^2/h$). The critical point is close to the expected semicircle although it is not at the top of the semicircle. In addition, the trace $\sigma_{xx}(\sigma_{xy})$ is also close to the expected semicircle on the right-hand side of the critical point, so features of $\Gamma(2)$ symmetry are observed. On the other hand, $\sigma_{xx}(\sigma_{xy})$ deviates from the semicircle on the left-hand side of the critical point at the same temperature. Therefore, the modular symmetry can be reduced to $\Gamma(2)$, and the exactness of the semicircle law can be relaxed under the paired Landau bands. As mentioned above, the deviations of the critical conductivities may be due to the generalization of particle-hole transformation, and the deviation of the semicircle law is expected when such a transformation has to be modified under a small spin gap. Since we observed these effects in both the p-type Si/SiGe and n-type GaAs systems, modifications to the particle-hole transformation are important either when the spin splitting is much smaller than the cyclotron gap, or when these two gaps are comparable.

In sample C, the semicircle law holds well and the critical point is at the top of the semicircle in the 2-1 transition. Comparing with the 3-2 transition in the same sample, the 2-1 transition occurs at higher $B$ where the gaps inducing QH states are larger.

Therefore, the larger the gaps, the better the semicircle law and the universality of the critical conductivities. It can be seen in Fig. 3 that the critical points in the 2-0 and 1-0 I-QH transitions are both at the top of the semicircle in sample B, and the semicircle law is not followed well in the 2-1 transition. In sample A, however, there is a deviation on the critical point in the 1-0 I-QH transition. We note that such deviation in an I-QH transition may also be due to the inappropriate conversion from resistivities to conductivities.[2]

In this study, we only probe the transitions following the law of corresponding states [1]. For the IQHE in most realistic 2D systems, in fact, such a law is applicable only at high magnetic fields. In samples A and C, we only study the IQHE at high fields because of the invalidity of the law of corresponding states at low magnetic fields. On the other hand, the low-field 2-0 transition of sample B is consistent with such a law and hence we can compare it to other transitions.

In conclusion, we studied effects of Zeeman splitting in the IQHE. Under the paired Landau bands arising from Zeeman spin splitting, the exactness of semicircle law can be relaxed, and the modular symmetry can be reduced to $\Gamma(2)$. Our study indicates the importance of the modifications to the particle-hole transformation either when the spin gap is much smaller than the cyclotron gap, or when these two gaps are comparable.

This work is supported by the National Science Council and the Ministry of Education of the Republic of China. D. R. Hang acknowledges support from the NSC, Taiwan (grant no: NSC 93-2112-M-110-007). Gil-Ho Kim was supported by National R&D Project for Nano Science and Technology (contract no M1-0212-04-0003) of MOST. The work undertaken in Cambridge was funded by the EPSRC, United Kingdom. Two of the authors (D. R. H. and C. F. H.) thank Drs. B. P. Dolan and C. P. Burgess for their valuable discussions.

| Sample | System | Transitions |
|---|---|---|
| A[3] | 2D GaAs electron system | 2-1, 1-0 |
| B[6] | 2D GaAs electron system with InAs dots | 2-0, 2-1, 1-0 |
| C[7] | 2D SiGe hole system | 3-2, 2-1 |

Table I: The systems and transitions under study.

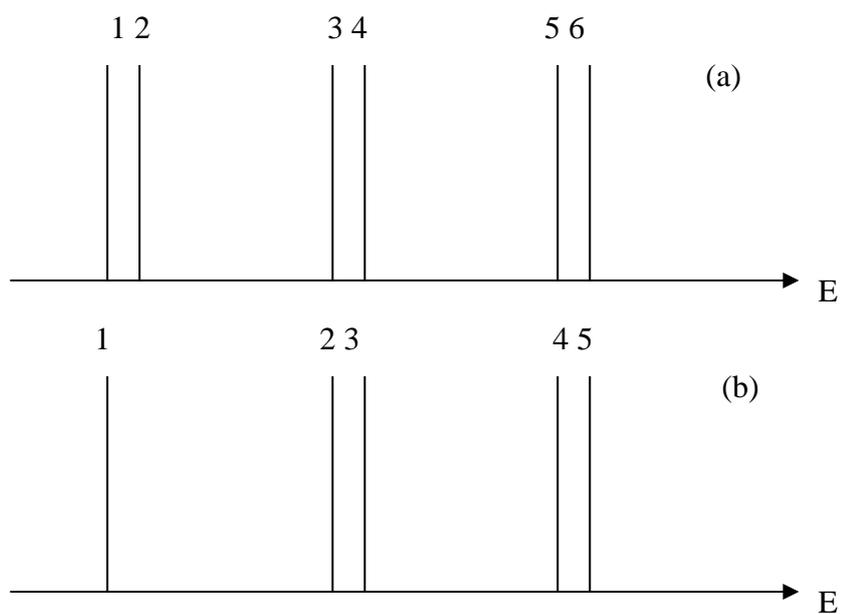

Fig. 1: (a) The spin configuration for sample A and B. (b) The spin configuration for sample C.

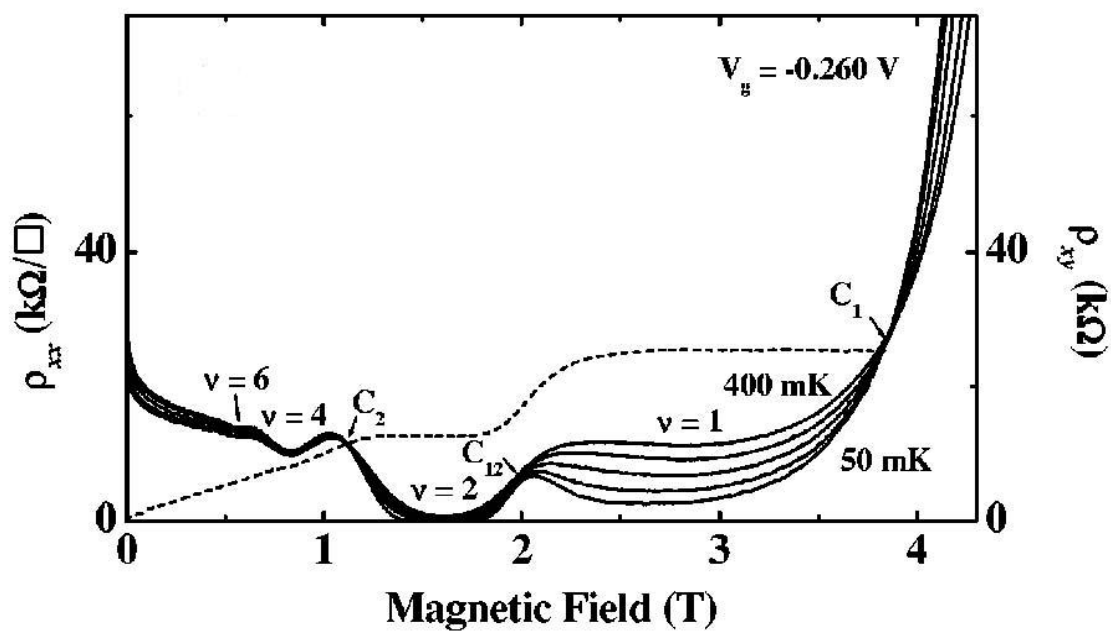

Fig.2 The traces of the longitudinal and Hall resistivities of sample B. [6]

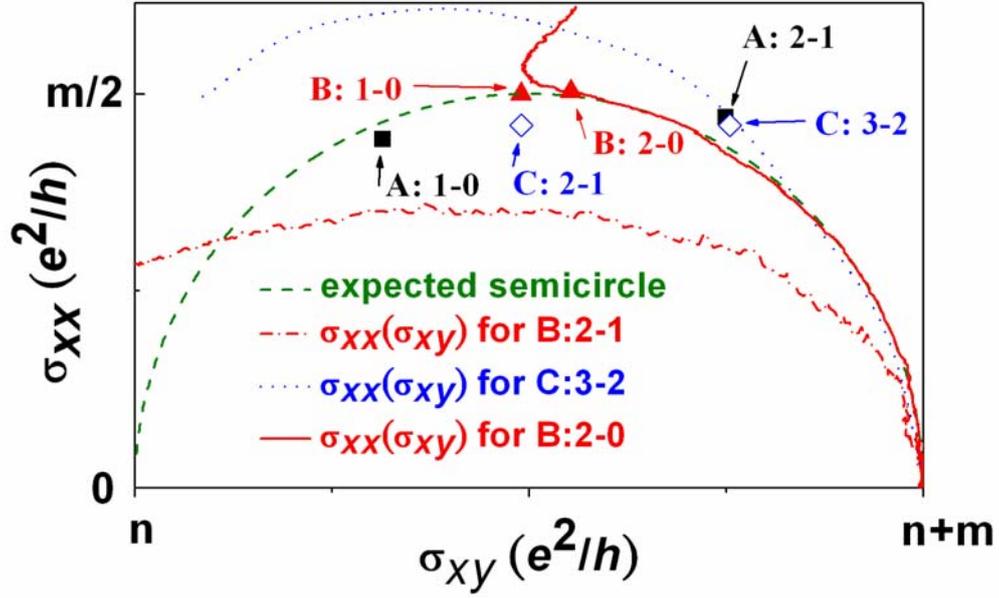

Fig. 3 The trace of $\sigma_{xx}(\sigma_{xy})$ and critical point. Here n = 0, 0, 1, and 2 for the 1-0, 2-0, 2-1, and 3-2 transitions, respectively. In addition, m = 2 for the 2-0 transition and m = 1 for all the other transitions. Each symbol corresponds to a critical point.

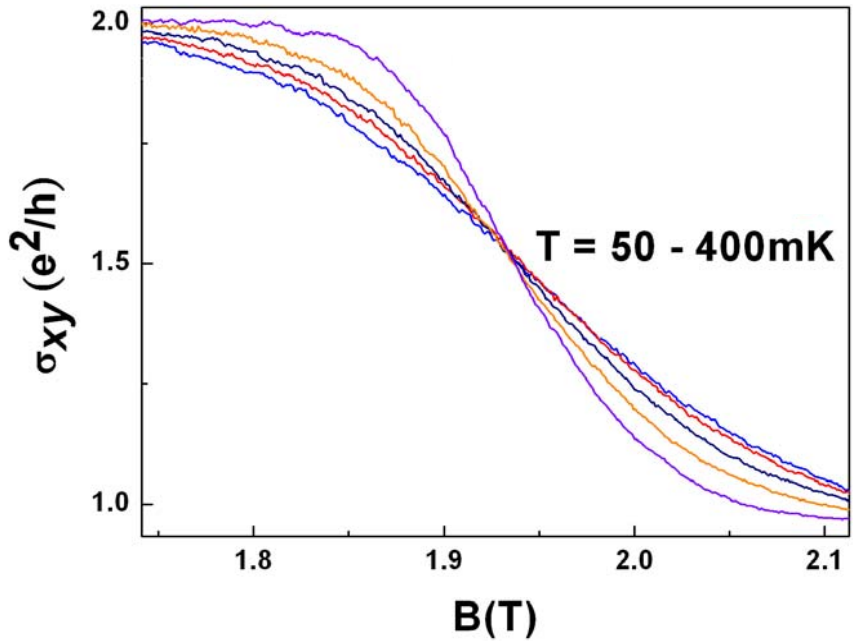

Fig. 4 The traces of the Hall conductivity in the 2-1 transition in sample B.